\documentstyle[12pt]{article}

\title{Individuality as an illusion}
\author{Adonai S. Sant'Anna\thanks{Permanent address:
Department of Mathematics, Federal University of Paran\'a, P. O.
Box 019081, Curitiba, PR, 81531-990. E-mail: adonai@ufpr.br.}}

\date{Department of Philosophy\\University of South Carolina\\Columbia, SC, 29208}
\begin{document}

\newtheorem{definicao}{Definition}
\newtheorem{teorema}{Theorem}
\newtheorem{lema}{Lemma}
\newtheorem{corolario}{Corollary}
\newtheorem{proposicao}{Proposition}
\newtheorem{axioma}{Axiom}
\newtheorem{observacao}{Observation}
\maketitle


\begin{abstract}

Elementary particles in quantum mechanics (QM) are
indistinguishable when sharing the same intrinsic properties and
the same quantum state. So, we can consider quantum particles as
non-individuals, although non-individuality is usually considered
as a consequence of the formalism of QM, since the entanglement of
states forbids any labelling process. We show how to consider
non-individuality as one of the basic principles of QM, instead of
a logical consequence. The advantages of our framework are
discussed as well. We also show that even in classical particle
mechanics it is possible to consider the existence of
non-individual particles. One of our main contributions is to show
how to derive the apparent individuality of classical particles
from the assumption that all physical objects are non-individuals.

\end{abstract}


\section{Introduction}

The issues of non-individuality in quantum physics have motivated
many research projects. See, for example, the references in
\cite{French-04}.

Elementary particles in quantum mechanics (QM) that share the same
set of state-independent (intrinsic) properties are sometimes said
to be {\em indistinguishable\/}. It is not possible, e.g., to keep
track of individual particles in order to distinguish among them
when they share the same physical properties. In other words, it
is not possible, in principle, to label quantum particles. This
non-individuality plays a very important role in quantum mechanics
\cite{Sakurai-94}; it is important in the derivation of quantum
statistics and in the analysis of the wave-function of atoms, for
example.

On the possibility that collections of such indistinguishable
entities should not be considered as sets in the usual sense, Yu.
Manin \cite{Manin-76} has proposed the search for axioms which
should allow to deal with indiscernible objects. As he said,

\begin{quote}
I would like to point out that it [standard set theory] is rather
an extrapolation of common-place physics, where we can distinguish
things, count them, put them in some order, etc. New quantum
physics has shown us models of entities with quite different
behavior. Even {\em sets\/} of photons in a looking-glass box, or
of electrons in a nickel piece are much less Cantorian than the
{\em sets\/} of grains of sand.
\end{quote}

We are using the philosophical jargon in saying that
`indistinguishable' objects are objects that share their
properties, while `identical' objects means `the very same
object'.

One manner to cope with the problem of non-individuality in
quantum physics is by means of quasi-set theory
\cite{Krause-92,Krause-99,Sant'Anna-00}, which is an extension of
Zermelo-Fraenkel set theory, that allows us to talk about certain
indistinguishable objects that are not identical. Such
indistinguishable objects are termed as non-individuals. In
quasi-set theory identity does not apply to all objects. There are
some situations in quasi-set theory where the sequence of symbols
$x = y$ is not a well-formed formula, i.e., it is meaningless. A
weaker equivalence relation called ``indistinguishability'' is an
extension of identity in the sense that it allows the existence of
{\em two\/} objects that are indistinguishable. In standard
mathematics, there is no sense in saying that two objects are
identical. If $x = y$, then we are talking about one single object
with two labels, namely, $x$ and $y$. But it is meaningful to say
that {\em two\/} objects are indistinguishable in quasi-set
theory.

Quasi-set theory has found some applications in quantum physics.
It has been used for an authentic proof of the quantum
distributions \cite{Krause-99}. By ``authentic proof'' we mean a
proof where elementary quantum particles are really considered as
non-individuals from the formal point of view. If the physicist
says that some particles are indistinguishable (in a sense) and
he/she still uses standard mathematics in order to cope with these
particles, then something does not seem to be sound, since
standard mathematics is based on the concept of individuality, in
the sense that it is grounded on the very notion of identity. It
was also proved \cite{Sant'Anna-00} that even non-individuals may
present a classical distribution like Maxwell-Boltzmann's. That is
another way to say that a Maxwell-Boltzmann distribution in an
ensemble of particles does not entail any ontological character
concerning such particles, as it was previously advocated by Nick
Huggett \cite{Huggett-99}. Besides, in \cite{Krause-99} the
authors also introduced the quasi-set-theoretical version of the
wave-function of the atom of Helium, which is a well known example
where indistinguishability plays an important role.

It is worth to remark that some authors like P. Pesic
\cite{Pesic-03} have advocated the idea that the non individuality
of elementary quantum particles should be considered as the
starting point on the foundations of quantum theory, instead of a
consequence of other fundamental principles. The usual way in QM
is to consider that the non-individuality of elementary particles
is a consequence of the standard formalism of QM, since the
entanglement of wave-functions would not allow any labeling
process to identify particles.

In this paper we offer a perspective quite different from the
standard approach. We take into consideration Pesic's ideas and
try a way where non-individuality is one of the basic assumptions
of quantum theory.

Our mathematical framework allows us to discuss another point. We
show that a similar sort of non-individuality may happen even in
classical particle mechanics. We believe that this may be useful
in some realistic interpretations of quantum mechanics, like
Bohmian mechanics.

\section{Non-individuals in quantum mechanics}

Consider two white clouds in the sky, separated by ordinary space.
Consider also that after some minutes these two clouds mix
together. Now, instead of two clouds, we have just one cloud.

Electrons and quantum elementary particles in general behave
differently. The state of one electron is like a cloud,
represented by its wave-function. But when the wave-functions of
two electrons mix together (get entangled) we still have two
electrons, despite the fact that we cannot distinguish the
particles. When two quantum states get entangled, we have only one
resulting quantum state. But the issue is that this new entangled
state is somehow associated to {\em two\/} particles.

Let us consider a very simple example from the literature, where
indistinguishability plays its role.

The original Einstein-Podolsky-Rosen (EPR) {\em Gedanken\/}
experiment deals with measurements of position and momentum in a
two-particles system. Here we use a composite 1/2-spin system
introduced by D. Bohm and inspired on EPR. We refer to this kind
of experimental setup as Einstein-Podolsky-Rosen-Bohm (EPRB)
experiment. Our discussion on this topic is essentially based on
Sakurai's textbook \cite{Sakurai-94}.

It is well known that the state ket of a two-electron system in a
spin-singlet state can be described by:

\begin{equation}
\Psi = \frac{1}{\sqrt{2}}(|z+;z-\rangle -
|z-;z+\rangle),\label{psi}
\end{equation}

\noindent where $z$ is an arbitrary quantization direction. The
physical interpretation of $|z+;z-\rangle$ and $|z-;z+\rangle$
depends on the measurement process. The component $|z+;z-\rangle$
means that electron 1 is in the spin up state and electron 2 is in
the spin down state, while $|z-;z+\rangle$ means that electron 1
is in the spin down state and electron 2 is in the spin up state.
Sometimes it is said that the $\Psi$ state is an entanglement of
two quantum states. These two entangled states correspond to the
two possible configurations after the spin measurement.

We may also rewrite equation \ref{psi} as:

\begin{equation}
\Psi_2 = \frac{1}{\sqrt{2}}(|z+;z-\rangle -
|z-;z+\rangle),\label{psi2}
\end{equation}

\noindent where the index corresponds to the cardinality of
a collection. The $\Psi_2$ state is associated to a collection
whose cardinality is 2. The main point is
that the elements of the two-elements collection are all quantum
particles of the same kind (since they are indistinguishable).
Nevertheless, after measurement, these indistinguishable
micro-atoms collapse to distinguishable macro-atoms $a$ and $b$,
i.e., they are individuals that can be identified (labelled) by
their quantum states, which are no longer entangled.

We cannot associate non-individuals of the two-particles
collection to different quantum states, since we are talking about
{\em indistinguishable\/} particles. But we can associate the
whole two-particle collection to the quantum state $\Psi$, as
suggested in equation (\ref{psi2}). And we certainly can associate
each one-particle collection to different quantum states, since
the after measurement particles are now distinguishable.

But the question is: how can we associate {\em two\/}
indistinguishable particles to one single quantum state? The
so-called entangled state has no information at all concerning the
number of associated particles. The entangled state is nothing but
a vector. A vector in the Hilbert space has no information about
the number of particles associated to it. So, how can we get the
information about the number of particles? One manner to answer
this question is by means of quasi-set theory.

\section{Quasi-sets}

This section is strongly based on other works
\cite{Krause-92,Krause-99,Sant'Anna-00}. We use standard logical
notation for first-order theories without identity
\cite{Mendelson-97}.

It is important to remark that, in contrast to the notions of set
and quasi-set, the term ``collection'' has an intuitive meaning in
this paper.

Quasi-set theory ${\cal Q}$ is based on Zermelo-Fraenkel-like
axioms and allows the presence of two sorts of atoms ({\it
Urelemente\/}), termed $m$-atoms (micro-atoms) and $M$-atoms
(macro-atoms). Concerning the $m$-atoms, a weaker `relation of
indistinguishability' (denoted by the symbol $\equiv$), is used
instead of identity, and it is postulated that $\equiv$ has the
properties of an equivalence relation. The predicate of equality
cannot be applied to the $m$-atoms, since no expression of the
form $x = y$ is a formula if $x$ or $y$ denote $m$-atoms. Hence,
there is a precise sense in saying that $m$-atoms can be
indistinguishable without being identical.

The universe of ${\cal Q}$ is composed by $m$-atoms, $M$-atoms and
{\it quasi-sets\/}. The axiomatization is adapted from that of ZFU
(Zermelo-Fraenkel with {\it Urelemente\/}), and when we restrict
the theory to the case which does not consider $m$-atoms,
quasi-set theory is essentially equivalent to ZFU, and the
corresponding quasi-sets can then be termed `sets' (similarly, if
also the $M$-atoms are ruled out, the theory collapses into ZFC).
The $M$-atoms play the same role of the {\it Urelemente\/} in ZFU.

In all that follows, $\exists_Q$ and $\forall_Q$ are the
quantifiers relativized to quasi-sets. That is, $Q(x)$ reads as
`$x$ is a quasi-set'.

In order to preserve the concept of identity for the
`well-behaved' objects, an {\it Extensional Equality\/} is defined
for those entities which are not $m$-atoms on the following
grounds: for all $x$ and $y$, if they are not $m$-atoms, then $$x
=_{E} y := \forall z ( z \in x \Leftrightarrow z \in y ) \vee
(M(x) \wedge M(y) \wedge x \equiv y).$$

It is possible to prove that $=_{E}$ has all the properties of
classical identity in a first order theory and so these properties
hold regarding $M$-atoms and `sets'. This happens because one of
the axioms of quasi-set theory says that the axiom of
substitutivity of standard identity holds only for extensional
equality. Concerning the more general relationship of
indistinguishability nothing else is said. In symbols, the first
axioms of ${\cal Q}$ are:

\begin{itemize}

\item $\forall x (x \equiv x)$,

\item $\forall x \forall y (x \equiv y \Rightarrow y \equiv x)$,
and

\item $\forall x \forall y \forall z (x \equiv y \wedge y \equiv z
\Rightarrow x \equiv z)$.

\end{itemize}

And the fourth axiom says that

\begin{itemize}

\item $\forall x \forall y (x =_{E} y \Rightarrow (A(x,x)
\Rightarrow A(x,y)))$, with the usual syntactic restrictions on
the occurrences of variables in the formula $A$.

\end{itemize}

In this text, all references to `$=$' (in quasi-set theory) stand
for `$=_E$', and similarly `$\leq$' and `$\geq$' stand,
respectively, for `$\leq_E$' and `$\geq_E$'. Among the specific
axioms of ${\cal Q}$, few of them deserve a more detailed
explanation. The other axioms are adapted from ZFU.

For instance, to form certain elementary quasi-sets, such as those
containing `two' objects, we cannot use something like the usual
`pair axiom', since its standard formulation assumes identity; we
use the weak relation of indistinguishability instead:

\begin{quote}

The `Weak-Pair' Axiom - For all $x$ and $y$, there exists a
quasi-set whose elements are the indistinguishable objects from
either $x$ or $y$. In symbols,

$$\forall x \forall y \exists_{Q} z \forall t (t \in z
\Leftrightarrow t \equiv x \vee t \equiv y).$$

\end{quote}

Such a quasi-set is denoted by $[x, y]$ and, when $x \equiv y$, we
have $[x]$, by definition. We remark that this quasi-set {\it
cannot\/} be regarded as the `singleton' of $x$, since its
elements are {\it all\/} the objects indistinguishable from $x$,
so its `cardinality' (see below) may be greater than $1$. A
concept of {\it strong singleton\/}, which plays a crucial role in
the applications of quasi-set theory, may be defined.

In ${\cal Q}$ we also assume a Separation Schema, which
intuitively says that from a quasi-set $x$ and a formula
$\alpha(t)$, we obtain a sub-quasi-set of $x$ denoted by $$[t\in x
: \alpha(t)].$$

We use the standard notation with `$\{$' and `$\}$' instead of
`$[$' and `$]$' only in the case where the quasi-set is a {\it
set\/}.

It is intuitive that the concept of {\it function\/} cannot also
be defined in the standard way, so a weaker concept of {\it
quasi-function\/} was introduced, which maps collections of
indistinguishable objects into collections of indistinguishable
objects; when there are no $m$-atoms involved, the concept is
reduced to that of function as usually understood. Relations (or
{\em quasi-relations\/}), however, can be defined in the usual
way, although no order relation can be defined on a quasi-set of
indistinguishable $m$-atoms, since partial and total orders
require antisymmetry, which cannot be stated without identity.
Asymmetry also cannot be supposed, for if $x \equiv y$, then for
every relation $R$ such that $\langle x, y \rangle \in R$, it
follows that $\langle x, y \rangle =_{E} [[x]] =_{E} \langle y, x
\rangle \in R$, by force of the axioms of ${\cal Q}$.

It is possible to define a translation from the language of ZFU
into the language of ${\cal Q}$ in such a way that we can obtain a
`copy' of ZFU in ${\cal Q}$. In this copy, all the usual
mathematical concepts (like those of cardinal, ordinal, etc.) can
be defined; the `sets' (actually, the `${\cal Q}$-sets' which are
`copies' of the ZFU-sets) turn out to be those quasi-sets whose
transitive closure (this concept is like the usual one) does not
contain $m$-atoms.

Although some authors like Weyl \cite{Weyl-49} sustain that
(concerning cardinals and ordinals) ``the concept of ordinal is
the primary one'', quantum mechanics seems to present strong
arguments for questioning this thesis, and the idea of presenting
collections which have a cardinal but not an ordinal is one of the
most basic and important assumptions of quasi-set theory.

The concept of {\it quasi-cardinal\/} is taken as primitive in
${\cal Q}$, subject to certain axioms that permit us to operate
with quasi-cardinals in a similar way to that of cardinals in
standard set theories. Among the axioms for quasi-cardinality, we
mention those below, but first we recall that in ${\cal Q}$,
$qc(x)$ stands for the `quasi-cardinal' of the quasi-set $x$,
while $Z(x)$ says that $x$ is a {\it set\/} (in ${\cal Q}$).
Furthermore, $Cd(x)$ and $card(x)$ mean `$x$ is a cardinal' and
`the cardinal of $x$', respectively, defined as usual in the
`copy' of ZFU.

\begin{quote}

Quasi-cardinality - Every quasi-set has an unique quasi-cardinal
which is a cardinal (as defined in the `ZFU-part' of the theory)
and, if the quasi-set is in particular a set, then this
quasi-cardinal is its cardinal {\em stricto sensu}:

$$\forall_{Q} x \exists_{Q} ! y (Cd(y) \wedge y =_{E} qc(x) \wedge
(Z(x) \Rightarrow y =_{E} card(x))).$$

\end{quote}

From the fact that $\emptyset$ is a set, it follows that its
quasi-cardinality is 0 (zero).

${\cal Q}$ still encompasses an axiom which says that if the
quasi-cardinal of a quasi-set $x$ is $\alpha$, then for every
quasi-cardinal $\beta \leq \alpha$, there is a sub-quasi-set of
$x$ whose quasi-cardinal is $\beta$, where the concept of {\it
sub-quasi-set\/} is like the usual one. In symbols,

\begin{quote}

The quasi-cardinals of sub-quasi-sets - $$\forall_{Q} x (qc(x)
=_{E} \alpha \Rightarrow \forall \beta (\beta \leq_{E} \alpha
\Rightarrow \exists_{Q} y (y \subseteq x \wedge qc(y) =_{E}
\beta)).$$

\end{quote}

Another axiom states that

\begin{quote}

The quasi-cardinal of the power quasi-set -
$$\forall_{Q} x (qc({\cal P}(x)) =_{E} 2^{qc(x)}).$$

\noindent where $2^{qc(x)}$ has its usual meaning.

\end{quote}

These last axioms allow us to talk about the quantity of elements
of a quasi-set, although we cannot count its elements in many
situations.

As remarked above, in ${\cal Q}$ there may exist quasi-sets whose
elements are $m$-atoms only, called `pure' quasi-sets.
Furthermore, it may be the case that the $m$-atoms of a pure
quasi-set $x$ are indistinguishable from one another. In this
case, the axiomatization provides the grounds for saying that
nothing in the theory can distinguish among the elements of $x$.
But, in this case, one could ask what it is that sustains the idea
that there is more than one entity in $x$. The answer is obtained
through the above mentioned axioms (among others, of course).
Since the quasi-cardinal of the power quasi-set of $x$ has
quasi-cardinal $2^{qc(x)}$, then if $qc(x) = \alpha$, for every
quasi-cardinal $\beta \leq \alpha$ there exists a sub-quasi-set $y
\subseteq x$ such that $qc(y) = \beta$, according to the axiom
about the quasi-cardinality of the sub-quasi-sets. Thus, if $qc(x)
= \alpha \not= 0$, the axiomatization does not forbid the
existence of $\alpha$ sub-quasi-sets of $x$ which can be regarded
as `singletons'.

Of course the theory cannot prove that these `unitary'
sub-quasi-sets (supposing now that $qc(x) \geq 2$) are distinct,
since we have no way of `identifying' their elements, but
quasi-set theory is compatible with this idea. In other words, it
is consistent with ${\cal Q}$ to advocate that $x$ has $\alpha$
elements, which may be regarded as absolutely indistinguishable
objects. Since the elements of $x$ may share the relation
$\equiv$, they may be further understood as belonging to the same
`equivalence class' but in such a way that we cannot assert either
that they are identical or that they are distinct from one
another.

The collections $x$ and $y$ are defined as {\it similar\/}
quasi-sets (in symbols, $Sim(x,y)$) if the elements of one of them
are indistinguishable from the elements of the other one, that is,
$Sim(x,y)$ if and only if $\forall z \forall t (z \in x \wedge t
\in y \Rightarrow z \equiv t)$. Furthermore, $x$ and $y$ are {\it
Q-Similar\/} ($QSim(x,y)$) if and only if they are similar and
have the same quasi-cardinality. Then, since the quotient
quasi-set $x/_{\equiv}$ may be regarded as a collection of
equivalence classes of indistinguishable objects, then the `weak'
axiom of extensionality is:

\begin{quote}

Weak Extensionality -
\begin{eqnarray}
\forall_{Q} x \forall_{Q} y (\forall z (z \in x/_{\equiv}
\Rightarrow \exists t (t \in y/_{\equiv} \wedge \, QSim(z,t))
\wedge \forall t(t \in
y/_{\equiv} \Rightarrow\nonumber\\
\exists z (z \in  x/_{\equiv} \wedge \, QSim(t,z)))) \Rightarrow x
\equiv y)\nonumber
\end{eqnarray}

\end{quote}

In other words, this axiom says that those quasi-sets that have
the same quantity of elements of the same sort (in the sense that
they belong to the same equivalence class of indistinguishable
objects) are indistinguishable.

\begin{definicao}
A {\em strong singleton\/} of $x$ is a quasi-set $x'$ which
satisfies the following property:
$$x' \subseteq [x] \wedge qc(x') =_{E} 1$$
\end{definicao}

\begin{definicao}
A {\em $n$-singleton\/} of $x$ is a quasi-set $[x]_n$ which
satisfies the following property:
$$[x]_n \subseteq [x] \wedge qc([x]_n) =_{E} n$$
\end{definicao}

\section{Standard quantum mechanics}

In this section we introduce an axiomatic framework for standard
quantum mechanics. We have eight primitive concepts, namely, $P$,
$H$, $O$, $I_P$, $M$, $T$, $f$, and $Pr$. All of them are sets (ZF
set theory). $P$ is a set of particles, $T$ is a time interval,
$H$ is a Hilbert space, $O$ is a set of Hermitean operators
defined on $H$ and corresponding to observables, $I_P$ is a set of
intrinsic properties, like rest mass, absolute value of spin,
electric charge, etc., $M$ is a set of functions termed the
``measurement functions'', $f$ is a function which associates to
each particle at each instant of time an intrinsic property and a
quantum state, and $Pr$ is a function that has a restriction which
is a probability function.

\begin{definicao}
${\cal QP} = \langle P, T, H, O, I_P, M, f, Pr \rangle$ is a
quantum system if and only if the following axioms are satisfied:

\begin{description}

\item [QP1] $P$ is a non-empty and finite set.

\item [QP2] $T$ is a non-degenerate interval of real numbers.

\item [QP3] $H$ is a Hilbert space with a norm induced by its
inner product. All vectors of $H$ are complex functions whose
domain is $\Re^3\times T$ (space and time).

\item [QP4] $O$ is a set of hermitean operators on $H$, such that
the eigenvectors of each operator from $O$ form a basis of $H$.

\item [QP5] $I_P$ is a set of ordered $n$-tuples of real numbers.

\item [QP6] $M$ is a class of functions $M_{O_i}:H\to H$, where
each $O_i \in O$.

\item [QP7] $f:P\times T\to I_P\times H$ is a function.

\item [QP8] $Pr:H\times H\to\Re$ is a real valued function, such
that $Pr(u,v) = |\langle u|v\rangle |^2$, where $\langle
u|v\rangle$ is the inner product between $u$ and $v$.

\end{description}

\end{definicao}

\begin{definicao}

For each operator $O_i\in O$ there is a set $S_i$ of normalized
eigenvectors $s_r^i$ of $O_i$, where each eigenvector $s_r^i$ is
associated to an eigenvalue $\lambda_i$, and the possible values
of $r$ depend on the dimension of $H$.

\end{definicao}

\begin{description}{\em

\item [QP9] If $H$ is spanned by a base $S_i$ of normalized
eigenvectors of an hermitean operator $O_i\in O$, then for every
vector $u\in H$, $M_{O_i}(u)\in S_i$.

\item [QP10] Each function of $M$ is a random function, in the
sense that every function of $M$ is associated to a probability
function; if $v\in S_i$, then the probability that $M_{O_i}(u) =
v$ is given by $Pr(u,v)$.

\item [QP11] Every vector of $H$ obeys the Schr\"odinger
equation.}

\end{description}

Axiom {\bf QP1} says that we are dealing with finite systems.
Axiom {\bf QP2} says that time flows on a continuum interval. {\bf
QP3} and {\bf QP4} are part of the standard mathematical
background of QM. {\bf QP5} says that intrinsic properties are
given by real numbers. {\bf QP6} is the first axiom of this
axiomatic framework concerning measurements in QM. {\bf QP7} is a
very strategic axiom, since it relates intrinsic properties to
quantum state properties (given by the vectors of $H$) by means of
the concept of particle. In other words, the notion of particle
has the role of connecting quantum states to intrinsic properties.
Since quantum particles can share the same intrinsic properties
and the same quantum state, this is a very easy solution to the
problem of representing ensembles of multiple indistinguishable
quantum particles. Within this context, particles may be
physically indistinguishable (by means of their physical
properties), although they are individuals in the sense of
belonging to a ``Cantorian'' set, namely, the set $P$. If we try
do describe a particle by means of its intrinsic properties and
quantum states only, then indistinguishability entails identity,
which forbids us to talk about collections of multiple
indistinguishable quantum particles. The challenge in the next
section is to consider indistinguishability on a new level of
formalism, namely, on a level where indistinguishability is
considered at the formalism itself. {\bf QP8} describes a function
that in some cases corresponds to a probability function, as
described by axiom {\bf QP10}. {\bf QP9} and {\bf QP10} are the
remaining axioms describing the measurement process. The last
axiom is a standard assumption which describes the time evolution
of undisturbed systems.

It is a theorem that $Pr(u,M_{O_i}(u))$ is a real number between 0
and 1 for all $O_i\in O$. One interesting exercise would be the
description of the $\sigma$-algebra associated to an appropriate
restriction of $Pr$. But that is not a task for this paper.

\section{Non-individuals in quantum mechanics}

In this section we introduce an alternative axiomatic framework
for QM, inspired on the idea of considering quantum particles
truly indistinguishable even on the formal language of the
axiomatic framework. The first obvious advantage of this is that
we have a mathematical framework more faithful to the usual
interpretation of physical phenomena. Hence, we can mathematically
justify quantum distributions and other physical effects where
indistinguishability plays its role. Another epistemological
advantage is that a quasi-set-theoretical approach to quantum
mechanics can justify the expression ``indistinguishable
particles'' without the need for an abstract concept like $P$,
whose physical interpretation is quite difficult. In other words,
if we already have all physical characteristics of a particle,
given by the elements of $I_P$ and $H$, what is the physical
meaning of $P$ in the previous axiomatic framework? We believe
that our quasi-set-theoretical solution to the problem of
non-individuality in quantum mechanics is more elegant from the
point of view of the foundations of physics.

We have eight primitive concepts, namely, $[x]_n$, $T$, $H$, $O$,
$I_P$, $M$, $P$, and $Pr$. $[x]_n$ is a $n$-singleton, $T$ is a
time interval, $H$ is a Hilbert space, $O$ is a set of Hermitean
operators defined on $H$ and corresponding to observables, $I_P$
is a set of intrinsic properties, $M$ is a set of functions termed
the ``measurement functions'', and $Pr$ is a function that has a
restriction which is a probability function.

\begin{definicao}

${\cal QP} = \langle [x]_n, T, H, O, I_P, M, P, Pr \rangle$ is a
quasi-quantum system if and only if the next axioms are satisfied:

\begin{description}

\item [QQP1] $[x]_n$ is a non-empty and finite $n$-singleton whose
elements are micro-atoms (we are using quasi-set-theoretical
terminology).

\item [QQP2] $T$ is a non-degenerate interval of real numbers.

\item [QQP3] $H$ is a Hilbert space with a norm induced by its
inner product. All vectors of $H$ are complex functions whose
domain is $\Re^3\times T$.

\item [QQP4] $O$ is a set of hermitean operators on $H$.

\item [QQP5] $I_P$ is a set of ordered $n$-tuples of real numbers.

\item [QQP6] $M$ is a class of functions $M_{O_i}:H\to H$, where
each $O_i \in O$.

\item [QQP7] $P$ is a sub-quasi-set of $[x]_n\times I_P\times H$
whose quasi-cardinality is $n$.

\item [QQP8] $Pr:H\times H\to\Re$ is a real valued function, such
that $Pr(u,v) = |\langle u|v\rangle |^2$, where $\langle
u|v\rangle$ is the inner product between $u$ and $v$.

\end{description}

\end{definicao}

\begin{definicao}

For each operator $O_i\in O$ there is a set $S_i$ of normalized
eigenvectors $s_r^i$ of $O_i$, where each eigenvector is
associated to an eigenvalue $\lambda_i$. The range of values of
$r$ depends on the dimension of $H$.

\end{definicao}

\begin{description}{\em

\item [QQP9] If $H$ is spanned by a set $S_i$ of normalized
eigenvectors of an Hermitean operator $O_i\in O$, then for every
vector $u\in H$, $M_{O_i}(u)\in S_i$.

\item [QQP10] Each $M_{O_i}$ of $M$ is a random function; if $v\in
S_i$, then for all $u\in H$, the probability that $M_{O_i}(u) = v$
is given by $Pr(u,v)$.

\item [QQP11] Every vector $u$ of $H$ obeys the Schr\"odinger
equation.}

\end{description}

The main difference between this system and the previous one is on
the concept of quantum particle. In the quantum system a particle
is an abstract object that belongs to a set $P$. So, the formal
language has all the ways to individualize any particle. In the
present system a particle is an ordered triple, where the first
element is an object devoid of individuality (a micro-atom), the
second element is an intrinsic property, and the third element is
a quantum state. The physical meaning of the first element of the
ordered triple is stated in section 8.

In the meantime, we emphasize that our approach makes possible to
say that a quantum state is associated to an ensemble of
indistinguishable particles. This is possible due to the concept
of quasi-cardinality (see section 3). In other words, in quasi-set
theory it is possible the existence of a $N$-singleton

\begin{equation}
X=_E [\langle x,i_p,u\rangle ]_N,
\end{equation}

\noindent where $x$ is a micro-atom, $i_p$ is an ordered $m$-tuple
of intrinsic properties, $u$ is the quantum state associated to
all particles of $X$, and $N$ is the number of particles of $X$.

So, recalling equation (\ref{psi2}),

\begin{equation}
\Psi_2 = \frac{1}{\sqrt{2}}(|z+;z-\rangle - |z-;z+\rangle),
\end{equation}

\noindent we can interpret the index 2 as the quasi-cardinality of
an ensemble of quantum particles that share the quantum state
$\Psi$.

Now, we need to illustrate some ideas concerning the elimination
of individuality even in classical particle mechanics. But first,
let us recall what do we mean by classical particle mechanics.

\section{Classical particle mechanics}

Consider a very simple and well known mathematical framework for
classical particle mechanics, in the newtonian formalism,
introduced by McKinsey, Sugar, and Suppes \cite{McKinsey-53}. We
call this as McKinsey-Sugar-Suppes (MSS) system for classical
particle mechanics, or MSS system, for short.

MSS system has six primitive notions: $P$, $T$, $m$, ${\bf s}$,
${\bf f}$, and ${\bf g}$. $P$ and $T$ are sets, $m$ is a
real-valued unary function defined on $P$, ${\bf s}$ and ${\bf g}$
are vector-valued functions defined on the Cartesian product
$P\times T$, and ${\bf f}$ is a vector-valued function defined on
the Cartesian product $P\times P\times T$. Intuitively, $P$
corresponds to the set of particles and $T$ is to be physically
interpreted as a set of real numbers measuring elapsed times (in
terms of some unit of time, and measured from some origin of
time). $m(p)$ is to be interpreted as the numerical value of the
mass of $p\in P$. ${\bf s}_{p}(t)$, where $t\in T$, is a
$3$-dimensional vector which is to be physically interpreted as
the position of particle $p$ at instant $t$. ${\bf f}(p,q,t)$,
where $p$, $q\in P$, corresponds to the internal force that
particle $q$ exerts over $p$, at instant $t$. And finally, the
function ${\bf g}(p,t)$ is to be understood as the external force
acting on particle $p$ at instant $t$.

Next we present the axiomatic formulation for MSS system:

\begin{definicao}
${\cal CP} = \langle P,T,{\bf s},m,{\bf f},{\bf g}\rangle$ is an
MSS system if and only if the following axioms are satisfied:

\begin{description}

\item [P1] $P$ is a non-empty, finite set.

\item [P2] $T$ is an interval of real numbers.

\item [P3] If $p\in P$ and $t\in T$, then ${\bf s}_{p}(t)$ is a
$3$-dimensional vector (${\bf s}_p(t)\in\Re^3$) such that
$\frac{d^{2}{\bf s}_{p}(t)}{dt^{2}}$ exists.

\item [P4] If $p\in P$, then $m(p)$ is a positive real number.

\item [P5] If $p,q\in P$ and $t\in T$, then ${\bf f}(p,q,t) =
-{\bf f}(q,p,t)$.

\item [P6] If $p,q\in P$ and $t\in T$, then $[{\bf s}_{p}(t), {\bf
f}(p,q,t)] = -[{\bf s}_{q}(t), {\bf f}(q,p,t)]$.

\item [P7] If $p,q\in P$ and $t\in T$, then $m(p)\frac{d^{2}{\bf
s}_{p}(t)}{dt^{2}} = \sum_{q\in P}{\bf f}(p,q,t) + {\bf g}(p,t).$
\end{description}
\end{definicao}

The brackets [,] in axiom {\bf P6} denote external product.

Axiom {\bf P5} corresponds to a weak version of Newton's Third
Law: to every force there is always a counter-force. Axioms {\bf
P6} and {\bf P5}, correspond to the strong version of Newton's
Third Law. Axiom {\bf P6} establishes that the direction of force
and counter-force is the direction of the line defined by the
coordinates of particles $p$ and $q$.

Axiom {\bf P7} corresponds to Newton's Second Law.

\begin{definicao}

Let ${\cal P} = \langle P,T,{\bf s},m,{\bf f},{\bf g}\rangle$ be a
MSS system, let $P'$ be a non-empty subset of $P$, let ${\bf s}'$,
${\bf g}'$, and $m'$ be, respectively, the restrictions of
functions ${\bf s}$, ${\bf g}$, and $m$ with their first arguments
restricted to $P'$, and let ${\bf f}'$ be the restriction of ${\bf
f}$ with its first two arguments restricted to $P'$. Then ${\cal
P'} = \langle P',T,{\bf s}',m',{\bf f}',{\bf g}'\rangle$ is a
subsystem of ${\cal P}$ if $\forall p,q\in P'$ and $\forall t\in
T$,

\begin{equation}
m'(p)\frac{d^{2}{\bf s}'_{p}(t)}{dt^{2}} = \sum_{q\in P'}{\bf
f}'(p,q,t) + {\bf g}'(p,t).
\end{equation}\label{P7}

\end{definicao}

\begin{teorema}
Every subsystem of a MSS system is again a MSS system.
\end{teorema}

\begin{definicao}
Two MSS systems \[{\cal P} = \langle P,T,{\bf s},m,{\bf f},{\bf
g}\rangle\] and \[{\cal P'} = \langle P',T',{\bf s}',m',{\bf
f}',{\bf g}'\rangle\] are equivalent if and only if $P=P'$,
$T=T'$, ${\bf s}={\bf s}'$, and $m=m'$.
\end{definicao}

\begin{definicao}
A MSS system is isolated if and only if for every $p\in P$ and
$t\in T$, ${\bf g}(p,t) = \langle 0,0,0\rangle$.
\end{definicao}

\begin{teorema}
If \[{\cal P} = \langle P,T,{\bf s},m,{\bf f},{\bf g}\rangle\] and
\[{\cal P'} = \langle P',T',{\bf s}',m',{\bf f}',{\bf g}'\rangle\]
are two equivalent systems of particle mechanics, then for every
$p\in P$ and $t\in T$
\[\sum_{q\in P}{\bf f}(p,q,t) + {\bf g}(p,t) = \sum_{q\in P'}{\bf f}'(p,q,t) + {\bf g}'(p,t).\]\label{somaforcas}
\end{teorema}

The embedding theorem is the following:

\begin{teorema}
Every MSS system is equivalent to a subsystem of an isolated MSS
system.\label{Her}
\end{teorema}

The next theorem can easily be proved by Padoa's method:

\begin{teorema}
Mass and internal force are each independent of the remaining
primitive notions of MSS system, i.e., they cannot be defined by
means of either $P$, $T$, ${\bf s}$ or ${\bf g}$.
\end{teorema}

The next theorem is proved in \cite{daCosta-01}.

\begin{teorema}\label{definet}
Time is definable from the remaining primitive concepts of MSS
system.
\end{teorema}

There are some analysis in the literature concerning MSS system.
See, for example, \cite{Sant'Anna-96,Sant'Anna-01,Sant'Anna-03}.
According to S. Obradovic \cite{Obradovic-02}, MSS system was
``the first successful axiomatization in classical mechanics of a
material point''. Besides, this system seems to be very reasonable
at first sight.

Now it is important to settle some terminology for this paper. MSS
system and classical particle mechanics are not the same concept.
MSS system is a formal mathematical framework described in the
definition given above, which is inspired on the common sense of
what physicists understand as classical particle mechanics.
Classical particle mechanics is not a formal theory from the
logical point of view. Classical particle mechanics is a paradigm
in theoretical physics which is grounded on the newtonian view of
mechanics.

One of the main advantages of MSS system, compared to classical
particle mechanics, is that particles are not considered as
points. Particles in MSS system are abstract objects that are
associated to mass, position, speed, and forces. As it happens in
classical particle mechanics, in MSS system the size and the shape
of particles (physical objects) are irrelevant. This is very
important for the understanding of the next section.

\section{Non-individuals in classical particle mechanics}

Particles in MSS system are the elements of $P$. The only
intrinsic (state independent) property of each particle in MSS
system is mass. All other physical properties, like position,
velocity, acceleration, and forces, are state dependent. The state
of a particle is the ordered pair

$$\left< {\bf s}_p(t), \frac{d{\bf s}_p(t)}{dt}\right>,$$

\noindent for all instant of time $t\in T$.

Forces, for example, are dependent on $\frac{d{\bf s}_p(t)}{dt}$,
according to axiom {\bf P7}.

This means that in MSS system there may exist two or more
particles with the same intrinsic and state dependent physical
properties. MSS system does nor forbid this kind of situation. It
is legitimate to consider two particles $1$ and $2$ such that $m_1
= m_2$ and such that for a period of time $T'$ their trajectories
and velocities are identical. But in classical particle mechanics
and particularly in MSS system all particles are distinguishable.
So, this kind of situation may create some confusion. In many
classical books on quantum mechanics it is said that elementary
particles are indistinguishable when sharing the same intrinsic
properties and the same quantum state. But in MSS system and even
in classical particle mechanics there may exist particles that
share the same intrinsic properties and the same state. So, what
makes elementary quantum particles really indistinguishable? It
seems that the non-individuality of particles in quantum mechanics
has a deeper origin. It does not seem to be reasonable to explain
relations of indistinguishability by means of coincidence of
intrinsic properties and physical state.

A physicist could argue that the notion of physical state in
classical particle mechanics and the notion of quantum state in
quantum mechanics are not equivalent. So, any comparison between
quantum and classical behavior is not straightforward. Another way
to understand the issues of non-individuality in quantum mechanics
is by means of the concept of wave-function. Since wave-functions
can be entangled, indistinguishability would be just a consequence
of the standard formalism of quantum mechanics. Nevertheless, we
show in this paper that it is possible to consider some sort of
non-individuality postulate right at the start of an axiomatic
framework for quantum mechanics. So, non-individuality does not
need to be understood as a consequence from the postulates of
quantum theory. The problem of two physical objects occupying the
same position at the same time is discussed later, since it
demands some further assumptions.

In the next definition we propose an axiomatic framework for a
system based on classical particle mechanics, but with
non-individual particles. We hope that our axiomatic framework may
be useful for a better understanding of the meaning of
non-individuality even in quantum theory. The intuitive meaning of
the primitive concepts and postulates is given afterwards.

\begin{definicao}
$\langle [x]_n,T,S,M,P,{\bf f},{\bf g}\rangle$ is a quasi-MSS
system if the following axioms are satisfied:

\begin{description}

\item[QP1] $[x]_n$ is a $n$-singleton whose elements are micro-atoms.
The elements of $[x]_n$ are denoted by $x_\alpha$, $x_\beta$,
$x_\gamma$, $x_\delta$, and so on.

\item[QP2] $T$ is a non-degenerate interval of real numbers.

\item[QP3] $S$ is the family of all functions ${\bf
s}_a:T\to\Re^3$ that are twice differentiable. The images of
these functions are denoted by ${\bf s}_a(t)$, ${\bf s}_b(t)$,
${\bf s}_c(t)$, ${\bf s}_d(t)$, and so on.

\item[QP4] $M$ is the set of all positive real numbers. The
elements of $M$ are denoted by $m_r$, $m_s$, $m_u$, $m_v$, and so
on.

\item[QP5] $P$ is a sub-quasi-set of $[x]_n\times M\times S$ whose
quasi-cardinality is $n$.

\item[QP6] ${\bf f}$ is a quasi-function ${\bf f}:P\times P\times
T\to \Re^3$.

\item[QP7] ${\bf g}$ is a quasi-function ${\bf g}:P\times
T\to\Re^3$.

\item[QP8] For all elements of $P$ and for all $t\in T$,

$${\bf f}(x_\alpha,m_r,{\bf s}_a(t);x_\beta,m_s,{\bf s}_b(t);t) =
- {\bf f}(x_\gamma,m_s,{\bf s}_b(t);x_\delta,m_r,{\bf
s}_a(t);t).$$

\item[QP9] For all elements of $P$ and for all $t\in T$,

$$[{\bf s}_a(t), {\bf f}(x_\alpha,m_r,{\bf s}_a(t);x_\beta,m_s,{\bf s}_b(t);t)] =
- [{\bf s}_b(t), {\bf f}(x_\gamma,m_s,{\bf
s}_b(t);x_\delta,m_r,{\bf s}_a(t);t)].$$

\item[QP10] For all elements of $P$ and for all $t$ of $T$ we have

$$m_r\frac{d^2{\bf s}_a(t)}{dt^2} = \sum_P {\bf
f}(x_\alpha,m_r,{\bf s}_a(t);x_\beta,m_s,{\bf s}_b(t);t)+{\bf
g}(x_\beta,m_r,{\bf s}_a(t)),$$

\noindent where the summation is performed over $N$ terms if there
are $N$ indistinguishable elements on $P$, plus the remaining
distinguishable elements.

\end{description}

\end{definicao}

Particles are ordered triples. The first element of these triples
is a micro-atom from quasi-set theory. It is this first element
that is responsible for the lack of individuality of particles. In
other words, we consider that individuality is unnecessary for
particles, although indispensable for describing physical states
and intrinsic properties. The second element of the ordered triple
corresponds to the intrinsic property and the third element has
all necessary information to describe the physical state of each
particle, since the third element is not just position but a
position function with respect to time $t$. $T$ is the set of
instants of time and $M$ is the set of all possible values for
mass. Axioms {\bf QP8} to {\bf QP10} are very similar to axioms
{\bf P5} to {\bf P7} and have a similar physical meaning.

It is easy to see that two particles with the same mass and the
same state (position and speed) are indistinguishable, i.e.,

$$(x_\alpha,m_r,{\bf s}_a(t)) \equiv (x_\beta,m_r,{\bf s}_a(t)),$$

\noindent since $x_\alpha \equiv x_\beta$.

MSS and quasi-MSS systems are not equivalent. In MSS system it is
possible the existence of two particles that share the same mass
and the same state, but being associated to different internal and
external forces. This happen because the coincidence of physical
properties does not entail any true indistinguishability between
particles. In MSS system, particles do have individuality. In
quasi-MSS system this kind of situation is not possible. Since
internal and external forces are quasi-functions,
indistinguishable particles are supposed to be associated to
indistinguishable forces. But forces are real three-dimensional
vectors which may be described by standard mathematics. So,
indistinguishable forces are identical forces.

\section{The physical meaning of $[x]_n$}

In \cite{French-04} we find an interesting statement:

\begin{quote}

[T]hat a permutation of the particles is counted as giving a
different arrangement in classical statistical mechanics implies
that, although they are indistinguishable, such particles can be
regarded as individuals (indeed, Boltzmann himself made this
explicit in the first axiom of his 'Lectures on Mechanics'). Since
this individuality resides in something over and above the
intrinsic properties of the particles in terms of which they can
be regarded as indistinguishable, it has been called
'Transcendental Individuality' by Post [see \cite{Post-63}].

\end{quote}

In this paper we consider a very different idea which goes to an
opposite direction. We consider that the notion of individuality
of physical objects may be considered as an illusion, caused by
some physical laws and our interpretation of them. In other words,
all physical objects in nature may be devoid of individuality, no
matter if they are subject to the laws of the macroscopic world
(classical mechanics) or the microscopic world (quantum
mechanics). In this sense, all physical objects would have some
sort of `Transcendental Non-individuality', where the illusion of
individuality would be caused by other physical components and our
interaction with the world.

Consider, as an example, the quasi-MSS system. If we add a new
axiom that says that the only internal force between particles is
Newtonian gravitation

\begin{equation}
{\bf f}(x_\alpha,m_r,{\bf s}_a(t);x_\beta,m_s,{\bf s}_b(t);t) =
\gamma\frac{m_r m_s}{\|{\bf s}_a(t)-{\bf s}_b(t)\|^3}({\bf
s}_a(t)-{\bf s}_b(t)),
\end{equation}

\noindent where $\gamma$ is the universal gravitational constant,
then we will never have any two particles sharing the same state,
since any coincidence of position would entail an inconsistency,
namely, a singularity on the gravitational potential. In this
case, any particle, despite its lack of individuality, can be
distinguished by its physical state. This means that in a
classical world we do not need any kind of transcendental
individuality to justify the apparent individuality of physical
objects. In the case of continuum mechanics another intrinsic
property plays a very important role in the process of
individuation: the size of the objects. So, we do not need the
assumption of individuality for physical objects (either classical
or quantum), although we need individual measures of state and
intrinsic properties. Hence, the physical meaning of $[x]_n$ is
that it drops the unnecessary individuality of a physical object
which is associated to individual measures of state and
state-independent properties. So, let us not confuse physical
properties with physical objects like particles. Physical
properties do need individuality. But physical objects don't.

Since QM does not forbid the existence of particles sharing all
their physical properties, we have the impression that
non-individuality is an exclusivity of the quantum world. But that
does not need to be true.

We are not proving that the whole world is made of
non-individuals. But we are pointing out to the possibility that
this may be a fact.

Another interesting issue concerns some realistic interpretations
of QM. Bohmian mechanics, for example, is a realistic
interpretation for quantum mechanics, where particles have
definite trajectories and velocities. So, some sort of classical
behavior among particles is unavoidable. The non-individuality of
elementary particles in Bohmian mechanics has more to do with
experimental facts than with the mathematical formalism.
Nevertheless we may consider the non-individuality in Bohmian
mechanics as an intrinsic characteristic of its own formalism.
That would be a manner to explain quantum statistics into the
scope of Bohmian mechanics.

\section{Acknowledgements}

This paper was quite improved thanks to some discussions with
Ot\'avio Bueno, Newton C. A. da Costa, D\'ecio Krause, and Michael
Dickson.

I would like to thank Ot\'avio Bueno and Davis Baird for their
hospitality during my stay at the Department of Philosophy of the
University of South Carolina.

This work was partially supported by CAPES (Brazilian government
agency).

\end{document}